\documentclass[twocolumn,prb,showpacs,superscriptaddress]{revtex4-1}
\usepackage{graphicx,amssymb,amsmath,bm}
\usepackage{xcolor}
\usepackage{float}
\usepackage{epstopdf}
\usepackage[
colorlinks=true,
citecolor=blue,
setpagesize=false]{hyperref}

\begin{document}

\title{Spin-orbital excitons and their potential condensation in pentavalent iridates}

\author {Beom Hyun Kim}
\affiliation{Korea Institute for Advanced Study, Seoul 02455, South Korea}
\author {Dmitry V. Efremov}
\affiliation{Institute for Theoretical Solid State Physics, IFW Dresden, Helmholtzstrasse 20, 01069 Dresden, Germany}
\author {Jeroen van den Brink}
\affiliation{Institute for Theoretical Solid State Physics, IFW Dresden, Helmholtzstrasse 20, 01069 Dresden, Germany}
\date{\today}

\begin{abstract}
We investigate magnetic excitations in iridium insulators with pentavalent
Ir$^{5+}$ ($5d^4$) ions with strong spin-orbit coupling. We
obtain a microscopic model based on the local Ir$^{5+}$ multiplets
involving $J=0$ (singlet), $J=1$ (triplet), and $J=2$ (quintet)
spin-orbital states. We get effective interactions between these
multiplets on square and face-centered-cubic (fcc) structures of
magnetic ions in the layered-perovskites and the
double-perovkites, in particular Ba$_2$YIrO$_6$. Further, we
derive an effective spin-orbital Hamiltonian in terms of bond
bosons and explore possible instabilities towards magnetic and
quadrupole orderings.
Additionally, we study charge excitations  with help of the variational cluster perturbation
theory and calculate the electronic charge gap as a function of hopping and Coulomb interactions.
Based on both electronic and magnetic phase diagrams, we verify
the possibility of excitonic magnetism due to condensation of
spin-orbital excitons in Ir$^{5+}$ iridates.

\end{abstract}

  %75.25.Dk,

\maketitle

\section{Introduction}
Magnetism in $4d$ or $5d$ transition-metal compounds has been an intriguing topic in condensed
matter physics for the last decade because of the strong interplay of electronic correlation
effects and sizable spin-orbit coupling. In contrast to the $3d$ transition metals they
have extended valence orbitals so that the electron-electron (e-e)
interactions is nearly screened out and local magnetic moment formation is suppressed.
Therefore magnetism in these materials is rather the exception than the rule.
Still there is a substantial number of magnetic compounds, most of them containing
Tc, Ru, Os, or Ir~\cite{Rodriguez2011,Longo1968,Lam2002,Cao1998,Singh2010}. Among these the iridates
are arguably the most-studied at the moment
\cite{BJKim2008,BJKim2009,Jackeli2009,Chaloupka2010,Kim2012a,Kim2012b,BHKim2012,BHKim2014}.
Examples are (Sr/Ba)$_2$IrO$_4$ \cite{BJKim2008,BJKim2009,Okabe2011,Boseggia2013}, Sr$_3$Ir$_2$O$_7$
\cite{Fujiyama2012,Carter2013}, (Na/Li)$_2$IrO$_3$
\cite{Singh2012,Manni2014,Gretarsson2013,Clancy2018,Hermann2018,Ducatman2018}.
In these tetravalent iridates compounds, the strong spin-orbit coupling (SOC) entangles locally the spin
and orbital degrees of freedom and causes splitting of $t_{2g}$ levels into a higher energy Kramers
doublet $j_{\rm eff}=1/2$ (pseudospin 1/2) and two pairs of lower energy ones
$j_{\rm eff}=3/2$ \cite{Jackeli2009}. Since a hole resides in doublet, an effect pseudospin 1/2
moment is formed.
%With the $d^5$ configuration of Ir$^{4+}$ the $j_{\rm eff} =3/2$ orbitals are fully occupied and $j_{\rm eff}=1/2$ are half-occupied.
%%
%The resulting bandwidth of the   $j_{\rm eff}=1/2$ electrons is much smaller than expected for the $t_{2g}$ electrons without SOC, what effectively enhances the correlations and leads to a magnetic Mott insulator.

Pentavalent Ir$^{5+}$ iridates with their $d^4$ local electron configuration on the other hand
are expected to be non-magnetic. The two holes in the $t_{2g}$ shell have parallel spin due to the
Hund's rule coupling, thus giving rise to a $S=1$ state, while at the same time the orbital angular
momentum of the two holes corresponds to $L=1$. Spin and orbital momentum are antiparallel so that the
local ground multiplet of Ir$^{5+}$ is a non-magnetic $J=0$ singlet.
Indeed, for instance the post-perovskite iridate NaIrO$_3$ exhibits a paramagnetic %(PM)
insulating behavior~\cite{Bremholm2011,Du2013}. 
However recently,
a new theoretical concept for unusual magnetism of the $d^4$
configuration in the presence of strong SOC was introduced in
Ref. [\onlinecite{Khaliullin2013}]. The main idea is as follows.
Without SOC the local ground multiplet of $d^4$  is $^3T_{2g}$
($L=1,S=1$) and when the SOC is turned on, the multiplet  is split
into non-degenerate $J=0$, triply degenerate $J=1$, and quintuply
degenerate $J=2$ states (see Fig.~\ref{fig1}(a)).
The magnetic excitation from $J=0$ to $J=1$ can be viewed as
simultaneous annihilation of singlet and creation of triplet
bosons, so called triplon excitons.
Similar to quantum dimer models~\cite{Sachdev2011},  the
superexchange (inter-dimer) interaction between multiplets can
bring about a dispersion of the excitation and the condensation of
triplons for some value of SOC (intra-dimer interaction). The
dipole condensation induces uncompensated magnetic moments. This
scenario was applied to explain magnetism and magnetic excitation
in layered-perovskite Ca$_2$RuO$_4$~\cite{Akbari2014}. In these
cases the quintet ($J=2$) states are usually not considered in the
theory, since the local multiplet energy of the quintet is three
times higher than the local energy of the triplets.

Recently a well-formed  magnetic moment and a clear magnetic order
below 1.3 K have been reported in Sr$_2$YIrO$_6$, in which the
Ir-ions nominally should have a $5d^4$ electron
configuration~\cite{Cao2014}. It was suggested that the unexpected
magnetism of Ir$^{5+}$ ($5d^4$ local ion configuration) may appear
due to the strong electron structure renormalization by the
non-cubic crystal field. This work initiated the discussion
whether the $5d^4$ systems can be magnetic. It motivated the
investigation of its sister material, cubic Ba$_2$YIrO$_6$, but no
such magnetism was found \cite{Corredor2017,Dey2016,Fuchs2018}.
Further studies of the magnetism in Sr$_2$YIrO$_6$ by other groups
did not give also positive results so far~\cite{Corredor2017}.
{Moreover, it was suggested that the magnetism in
Sr$_2$YIrO$_6$ initially reported in Ref.~[\onlinecite{Cao2014}]
likely is not intrinsic, rather an effect of a weak chemical
disorder~\cite{Chen2017,Fuchs2018}. The results of the recent
resonant inelastic x-ray scattering experiments on
momentum-dependence of the spin-orbital excitations in both
Sr$_2$YIrO$_6$ and Ba$_2$YIrO$_6$ support these
conclusions~\cite{Kusch2018}. In the experiments, low dispersive
($<50$meV) triplet ($J=1$) and quintet ($J=2$) excitations
(excitons) were observed at about 370 and 650 meV, respectively.
From this perspective, it is rather unlikely that the excited
triplet states can condense to form a magnetic state which also is
supported by recent theoretical
studies~\cite{Pajskr2016,Gong2018}. Nevertheless, the question,
whether spin-orbital condensed phase can in principle occur in
$4d^4$ and $5d^4$ systems by tuning the lattice parameters with
pressure or strain, is still open.

Here we study the possibility of condensation of the spin-orbital
excitons  (both triplet and quintet) in pentavalent iridates with
a layered- and/or the double-perovskite structure. We derive an
effective model of these states and incorporate exciton
dispersion.  In addition, we explore the electronic charge
excitation with the help of the variational cluster perturbation
theory (CPT).  We show that the charge gap is much more sensitive
to the electron hopping than that of the spin-orbital excitons in
the double-perovskite structure. We find that already in the
regime when the Coulomb correlation ($U$) is of the order or less
than the SOC ($\lambda$), the charge gap can close, leading to the
metal-insulator transition.

\section{Multiplet structure and spin-orbital states}

In order to set up a realistic effective Hamiltonian,  we employ
the local  Ir$^{5+}$ multiplets obtained by means of  {\it ab
initio} quantum chemistry and reported in
Ref.~[\onlinecite{Kusch2018}]. Using the microscopic model we
obtain the magnetic interactions of the different spin-orbital
states on square and face-centered-cubic (fcc) lattices.

\subsection{Local multiplet structure of Ir$^{5+}$}

Our starting point is  following Hamiltonian \cite{BHKim2012,BHKim2014}:
\begin{widetext}
\begin{align}
\label{eq1}
H_l &=\sum_{\mu\sigma} \epsilon_{\mu} n_{\mu\sigma}+
   \lambda \sum_{\mu\nu\sigma\sigma^{\prime}}
   (\mathbf{l}\cdot\mathbf{s})_{\mu\sigma,\nu\sigma^{\prime}}
   c_{\mu\sigma}^{\dagger}c_{\nu\sigma^{\prime}} \nonumber
  + \frac{1}{2}\sum_{\sigma\sigma^{\prime}\mu\nu}
  U_{\mu\nu} c_{\mu\sigma}^{\dagger}c_{\nu\sigma^{\prime}}^{\dagger}
  c_{\nu\sigma^{\prime}}c_{\mu\sigma}
  + \frac{1}{2}\sum_{\substack{\sigma \sigma^{\prime} \\ \mu\ne\nu}}
  J_{\mu\nu}c_{\mu\sigma}^{\dagger}c_{\nu\sigma^{\prime}}^{\dagger}
  c_{\mu\sigma^{\prime}}c_{\nu\sigma}  \nonumber \\
  +&\frac{1}{2}\sum_{\substack{\sigma \\ \mu\ne\nu}}
  J_{\mu\nu}^{\prime} c_{\mu\sigma}^{\dagger}c_{\mu\bar{\sigma}}^{\dagger}
  c_{\nu\bar{\sigma}}c_{\nu\sigma},
\end{align}
\end{widetext}
where  $U_{\mu\mu}=U$, $U_{\mu\ne\nu}=U-2J_H$ is the onsite Coulomb interaction,
$J_{\mu\nu}=J_{\mu\nu}^{\prime}=J_H$ represents Hund's coupling and $\lambda$ is the strength of %the spin-orbit coupling
the SOC for single $t_{2g}$ orbital which is twice as large as
that of the effective SOC for $t_{2g}^4$ multiplets in 
Ref.~\onlinecite{Khaliullin2013}.
$\bar{\sigma}$ stands for the opposite sign of $\sigma$.
Only three $t_{2g}$ orbitals are taken into account in the model.
According to our results, %[\cite{BHKim2012,BHKim2014}],
the  energies of the $J=1$ and $J=2$ multiplets strongly depend on
the  $J_H$ and $\lambda$ values. In the $J_H >> \lambda$ limit,
the energy of $J=2$ level ($E_{J=2}=3/2\lambda$) is three times as
large as  the energy of $J=1$ ($E_{J=1}=1/2\lambda$). In the
opposite limit their energies tend to same value ($3/2\lambda$)
(see Fig.~\ref{fig1}(b))%~\cite{Lambda}
because of non-negligible
overlap between high energy multiplets with $S=0$. This overlap
leads also shifting down of the multiplet levels of $J=0$ and
$J=2$. In our further calculations  we  adopt  $J_H =0.5$ and
$\lambda=0.4$ eV, which give the energy levels consistent with
those of the quantum chemistry calculation in
Ref.~[\onlinecite{Kusch2018}].

%%%%%%%%%%%%%%%%%%%%%%%%%%%%%%%%%%%%%%%%%%%%%%%%%%%%%%%%%%%%%%%
%%\begin{table}[b]
%%\caption {
%%Theoretical energies of local multiplets of $d^4$.
%%The number inside round brackets refers to
%%the degeneracy of multiplets with given energy. Unit is eV.
%%We set $J_H=$0.5 eV and $\lambda=$0.4 eV for the microscopic model.
%%$E_s$, $E_t$, and $E_q$ refer to the energy of $J=0$, $J=1$, and $J=2$, respectively.
%%}
%%\label{level}
%%\begin{ruledtabular}
%%\begin{tabular}{ c  c  c }
%%  & CAS+SOC & Model \\
%%\hline
%% $\quad E_s \quad$  &  0.00(1) & 0.00(1) \\
%% $\quad E_t \quad$  &  0.31(3) & 0.33(3) \\
%% $\quad E_q \quad$  &  0.63(4),0.64(1)& 0.65(5) \\
%% $\quad E_q^{\prime} \quad$  &  1.56(1),1.57(2),1.70(2) & 1.64(5) \\
%% $\vdots$ & $\vdots$  & $\vdots$ \\
%%\end{tabular}
%%\end{ruledtabular}
%%\end{table}
%%%%%%%%%%%%%%%%%%%%%%%%%%%%%%%%%%%%%%%%%%%%%%%%%%%%%%%%%%%%%%%

%%%%%%%%%%%%%%%%%%%%%%%%%%%%%%%%%%%%%%%%%%%%%%%%%%%%%%%%%%%%%
\begin{figure}[t]
\centering
\includegraphics[width=\columnwidth]{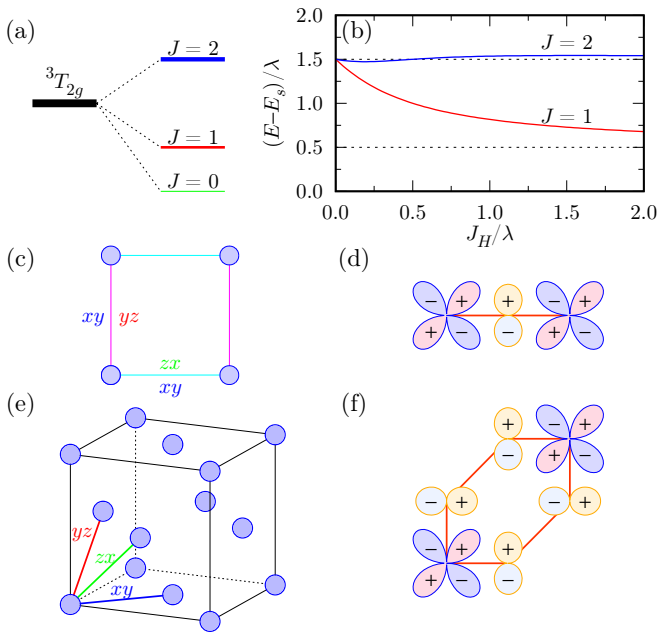}
\caption
{ (Color online)
   (a) Energy splitting of the local multiplets due to the strong spin-orbit coupling.
   (b) Excitation energies from the $J=0$ multiplet for the microscopic model
   as a function of Hund's coupling ($J_H$).
   $E_s$ refers to the ground state energy of the $J=0$ multiplet.
   Lattice structures of (c) corner-shared and (e) face-centered iridates,
   and their schematic orbital arrangements ((d) and (f)),% which
   which allow the hopping between adjacent iridiums.
   Orbital characters displayed in (c) and (e) represent dominant contributing
   orbitals between specific hopping directions.
}
\label{fig1}
\end{figure}
%%%%%%%%%%%%%%%%%%%%%%%%%%%%%%%%%%%%%%%%%%%%%%%%%%%%%%%%%%%%%

\subsection{Two Ir$^{5+}$ sites}

Now we study the magnetic exchange between nearest-neighboring
Ir-sites due to electron hopping described by the following
Hamiltonian:
\begin{equation}
H_t=\sum_{\alpha\beta\langle i,j \rangle} t_{\alpha\beta}
 c_{i\alpha}^{\dagger} c_{j\beta} +h.c..
\label{eq2}
\end{equation}
In the following we will consider two types of lattices: layered
perovskites and double perovskites. The dominant hoppings, found
for these two lattices in the limit of linear combination of the
atomic orbitals~\cite{Slater},  are illustrated in
Fig.~\ref{fig1}(d) and (f). In the layered perovskites the
Ir-O-Ir geometry is $180^\circ$-bond formed by corner-shared
octahedra as in Fig.~\ref{fig1}(d). 
%In this case the dominant hopping is between adjacent 
%$xy$ orbitals and $yz$ orbitals along the $x$-axis.
In this case the dominant hopping is 
between adjacent $xy$ orbitals 
and between $zx$ ($yz$) orbitals along the x-axis (y-axis).
Its value is $-V_{pd\pi}^2/\Delta$, where
$V_{pd\pi}$ and $\Delta$ are the $\pi$ bonding strength between
$p$ and $d$ orbital, and the charge transfer energy of oxygens,
respectively. In double perovskites the Ir geometry is
face-centered system as in Fig.~\ref{fig1}(f) and the dominant
hopping is between $xy$ orbitals in the $xy$-plane.  It can be
expressed as $-V_{pd\pi}^2(V_{pp\sigma}-V_{pp\pi})/\Delta^2$,
where $V_{pp\sigma}$ and $V_{pp\pi}$ are the $\sigma$ and $\pi$
bondings between $p$ orbitals. Although the hopping between $yz$
and $zx$ orbitals is also possible with the strength 
$-V_{pd\pi}^2 V_{pp\pi}/\Delta^2$,
it can be neglected in leading order, 
since $\left| V_{pp\pi}\right| \ll \left|V_{pp\sigma}\right|$.
%since $V_{pp\sigma} \ll V_{pp\pi}$. 
The obtained estimates for hopping
are consistent with the results from density functional theory
(DFT) calculations~\cite{Dey2016}. For these reasons we restrict
the consideration by two orbitals for the corner-shared system and
by one orbit for the face-centered lattices with the nearest
neighbors hopping  $t$ ($t<0$)  as it is illustrated in
Fig.~\ref{fig1}(c) and (e).

To  proceed further we construct the Hilbert space for a two-site
cluster that consists of all possible states whose configuration
is $d^4$-$d^4$, $d^3$-$d^5$, or $d^5$-$d^3$. Then we find the
eigenstate energies of the two-site cluster with help of the exact
diagonalization (ED) method. Figure~\ref{fig2} presents the energy
hierarchy and its partial density of states for $U=4.0$,
$J_H=0.5$, $\lambda=0.4$, and $t=-0.4$ eV.

%%%%%%%%%%%%%%%%%%%%%%%%%%%%%%%%%%%%%%%%%%%%%%%%%%%%%%%%%%%%%
\begin{figure}
\centering
\includegraphics[width=.9\columnwidth]{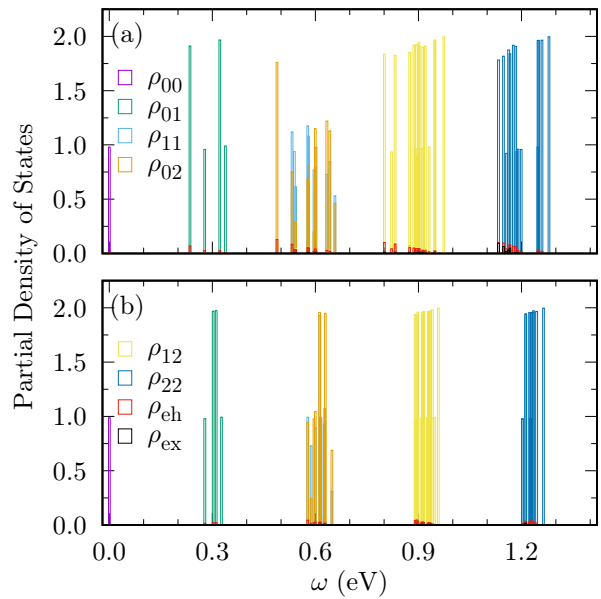}
\caption
{ (Color online)
   Partial density of states of lowest eighty-one states of the two-site cluster for (a)
   corner-shared and (b) face-centered hopping when $U$=4.0, $J_H$=0.5, $\lambda=0.4$,
   and $t=-0.4$ eV.
   $\rho_{j_1j_2}$ represent the partial density of states in which
   $J=j_1$ is stabilized in one site and $J=j_2$ in another site.
   $\rho_{eh}$ and $\rho_{ex}$ refer to the partial density of states with $d^3$-$d^5$ or
   $d^5$-$d^3$ configuration, and remain states with $d^4$-$d^4$ configuration, respectively.
   The total density of nondegenerate state is one.
}
\label{fig2}
\end{figure}
%%%%%%%%%%%%%%%%%%%%%%%%%%%%%%%%%%%%%%%%%%%%%%%%%%%%%%%%%%%%%

We introduce  the partial density of states  $\rho_{j_1j_2}$, in
which local multiplet state is one out of $J=j_1$ and $j_2$
states, $\rho_{eh}$ to be the partial density of states of
configuration is $d^3$-$d^5$ or $d^5$-$d^3$ (electron-hole (e-h)
excitation states), and $\rho_{ex}$ to be the remaining
multiplets. Since the e-h excitations appear above
$U-3J_H+\lambda=2.9$ eV, multiplet states below 1.4 eV
($>1.5\lambda$) can be attributed to the $L=1,S=1$ manifold.  We
find that $\rho_{eh}$ and $\rho_{ex}$ in the region are always
less than 10\% in these parameters. This implies that the magnetic
interactions between $L=1,~S=1$ multiplets are induced by the
virtual hopping like superexchange interactions which give rise to
the splitting of excitation multiplets. According to
Fig.~\ref{fig2}(a) and (b), the splittings in the corner-shared
system are almost twice as large as those in the face-centered
system. The corner-shared system has about $104$ meV of
single-triplet ($\rho_{01}$) and $169$ meV of single-quintet
($\rho_{02}$) splitting, whereas the splittings in the
face-centered system are $49$ meV and $71$ meV, correspondingly.
This agrees with the expectation that the corner-shared system
with two hopping channels has larger magnetic interactions than
the face-centered system with one hopping channel.

\section{Effective low-lying spin-orbital excitations}

\subsection{Effective excitonic Hamiltonian}
To derive an effective Hamiltonian for the  interaction between
different spin-orbital states,  we define $E_n$ and
$|\Psi_n\rangle$ be an eigenvalue and eigenstate of two-site
cluster, respectively. In the limit of $U-3J_H >> 2 \lambda$, the
effective interactions between $d^4$ multiplets ($L=1,S=1$) along
the $\delta$-direction can be extracted from the two-site
calculation as follows~\cite{BHKim2012,Winter2016}
\begin{align}
H_{\delta} &= \sum_{n=0}^{80} E_n \frac{ \mathcal{P}_{LS} |\Psi_n\rangle\langle\Psi_n|
\mathcal{P}_{LS} } {\langle\Psi_n|\mathcal{P}_{LS}^2|\Psi_n\rangle} \nonumber \\
 &= \sum_{j_1 j_2 j_1^{\prime} j_2^{\prime}} \sum_{mnm^{\prime}n^{\prime}}
 E_{mnm^{\prime}n^{\prime}}^{j_1 j_2 j_1^{\prime} j_2^{\prime}}(\delta)
 |T_{m}^{j_1} T_{n}^{j_2} \rangle \langle
  T_{m^{\prime}}^{j_1^{\prime}}T_{n^{\prime}}^{j_2^{\prime}}|
\label{Hcl}
\end{align}
where $T_{m}^{j}$ is the multiplet state with $J=j$ and $J_z=m$,
$\mathcal{P}_{LS}$ is the projection operator into the $L=1,S=1$
manifold, and $|\Psi_n\rangle$ is one of lowest $81$ states. Since
$J$-representation with $L=1$ and $S=1$  is very similar to a
quantum dimer model for a spin-$1$ system, we can introduce the
same bond-boson operators~\cite{Sachdev1990,Brenig2001} (see
appendix~\ref{append}) and derive the effective lattice
Hamiltonian in terms of these bosons by considering all possible
effective interactions incorporated in Eq.~\ref{Hcl}:
\begin{align}
& H_{\rm eff} =\sum_i \epsilon_s s_i^{\dagger}s_i + \sum_{i\alpha}
 \epsilon_t t^{\dagger}_{\alpha i} t_{\alpha i} +
 \sum_{i \mu} \epsilon_q q^{\dagger}_{\mu i} q_{\mu i}
 \nonumber \\
&~+\frac{1}{2}\sum_{i\delta \alpha\beta} \left( [h_{\delta}^{11}]_{\alpha\beta}
  s_i^{\dagger} t_{\alpha i_\delta}^{\dagger} t_{\beta i} s_{i_\delta}
 + [d_{\delta}^{11}]_{\alpha\beta}
  t_{\beta i}^{\dagger} t_{\alpha i_\delta}^{\dagger} s_i s_{i_\delta} +h.c. \right)
 \nonumber \\
&~+\frac{1}{2}\sum_{i\delta \mu\nu} \left(
 [h_{\delta}^{22}]_{\mu\nu}
 s_i^{\dagger} q_{\mu i_\delta}^{\dagger} q_{\nu i} s_{i_\delta}
 + [d_{\delta}^{22}]_{\mu\nu}
 q_{\nu i}^{\dagger} q_{\mu i_\delta}^{\dagger} s_i s_{i_\delta} +h.c. \right)
 \nonumber \\
&~+\frac{1}{2}\sum_{i\delta \alpha\mu} \left( [h_{\delta}^{12}]_{\alpha\mu}
  s_i^{\dagger} t_{\alpha i_\delta}^{\dagger} q_{\mu i} s_{i_\delta}
 + [d_{\delta}^{12}]_{\alpha\mu}
  q_{\mu i}^{\dagger} t_{\alpha i_\delta}^{\dagger} s_i s_{i_\delta} +h.c. \right)
 \nonumber \\
&~ +\frac{1}{2}\sum_{i\delta \mu\alpha} \left(
 [h_{\delta}^{21}]_{\mu\alpha}
 s_i^{\dagger} q_{\mu i_\delta}^{\dagger} t_{\alpha i} s_{i_\delta}
 + [d_{\delta}^{21}]_{\mu\alpha}
 t_{\alpha i}^{\dagger} q_{\mu i_\delta}^{\dagger} s_i s_{i_\delta} +h.c. \right)
 \nonumber \\
%&~ + \sum_{i\delta \alpha\beta} \left( h_{\alpha\beta}^\delta
%  s_i^{\dagger} t_{\alpha i+\delta}^{\dagger} t_{\beta i} s_{i+\delta}
% + d_{\alpha\beta}^\delta
%  t_{\beta i}^{\dagger} t_{\alpha i+\delta}^{\dagger} s_i s_{i+\delta} +h.c. \right)
% \nonumber \\
%&~ + \sum_{i\delta \mu\nu} \left(
%  \bar{h}_{\mu\nu}^\delta s_i^{\dagger} q_{\mu i+\delta}^{\dagger} q_{\nu i} s_{i+\delta}
% + \bar{d}_{\mu\nu}^\delta
%  q_{\nu i}^{\dagger} q_{\mu i+\delta}^{\dagger} s_i s_{i+\delta} +h.c. \right)
% \nonumber \\
&~ + \cdots,
\label{Heff}
\end{align}
where $2\epsilon_{s}=E_{0000}^{0000}(\delta)$,
$\epsilon_{t}~(\epsilon_{q}) +\epsilon_{s}= E_{\alpha 0 \alpha 0}^{1010}(\delta)
~(E_{\mu 0 \mu 0}^{2020}(\delta))$,
%$h_{\alpha\beta}^\delta~(\bar{h}_{\mu\nu}^\delta)=E_{0 \alpha\beta 0}^{0110}(\delta)
%~(E_{0 \mu\nu 0}^{0220}(\delta))$,
%and $d_{\alpha\beta}^\delta~(\bar{d}_{\mu\nu}^\delta)=
%E_{\alpha\beta 00}^{1100}(\delta) ~(E_{\mu\nu 00}^{2200}(\delta))$.
$[h_\delta^{jj'}]_{\tau\tau'}= E_{0 \tau\tau'0}^{0jj'0}(\delta)$,
and $[d_\delta^{jj'}]_{\tau\tau'}= E_{\tau\tau' 0 0}^{jj'00}(\delta)$.
$i_\delta$ refers to the neighboring site of $i$-th site along
the $\delta$ direction.

As it is shown in Appendix~\ref{inter_t}, the interaction between
triplet bosons ($\mathbf{h}_\delta^{11}$) is of the $XXZ$-type in
both corner-shared  and face-centered cases. Only the
$Z$-direction varies depending on the bonding type. In
corner-shared (face-centered) case, $Z$ will be $x$, $y$, and $z$
for the interaction along (in) the $x$- ($yz$-), $y$- ($zx$-), and
$z$-axis ($xy$-plane), respectively. The different nature of the
hopping, however, gives rise to a somewhat different
characteristics in these two cases: a Heisenberg interaction is
dominant in the corner-shared case, whereas the Ising interaction
prevails in the face-center case. In addition, the interaction
parameters between quintets ($\mathbf{h}_\delta^{22}$) show
different behaviors according to the bonding nature. The
interaction for  the $Z^2$-type quintet in the corner-shared case
gives minimum strength in contrast to the face-centered case where
it gives maximum strength. Because the interaction parameters
between triplets and quintets ($\mathbf{h}_\delta^{12}$ and
$\mathbf{h}_\delta^{21}$) have non-zero elements as shown in
table~\ref{TB3}, the coupling between triplets and quintets
emerges. $X$- and $Y$-type triplets can be coupled with $YZ$- and
$ZX$-type quintets, respectively, when the displacement of
neighboring sites is parallel to the $Z$-axis in the corner-shared
case and the $XY$-plane in the face-center case.

\subsection{Dispersion of spin-orbital excitations}
Note that only singlet bosons condense in the case of weak inter-site interactions.
To investigate the exciton dispersion and possible condensation in the triplet or quintet channel,
we treat Hamiltonian Eq.~\ref{Heff} in the mean-field approximation
(see the detail in appendix~\ref{meanH}) as following
\begin{align}
& H_{MF} \approx N\epsilon_s \nonumber \\
&~+ \sum_{\mathbf{k}} \left[
   \bm{\psi}^\dagger_\mathbf{k} \mathbf{h} (\mathbf{k})
   \bm{\psi}_\mathbf{k}+
   \frac{1}{2} \left(
   \bm{\psi}^\dagger_\mathbf{k} \mathbf{d} (\mathbf{k})
   \bm{\psi}^*_{-\mathbf{k}} +  h.c.
   \right) \right],
% + \sum_{\alpha \mathbf{k}} H_{\alpha}^t(\mathbf{k})
% + \sum_{\mu \mathbf{k}} H_{\mu}^q(\mathbf{k}) + \cdots \\
%& H_{\alpha}^t(\mathbf{k}) =  \epsilon_{\alpha\alpha} (\mathbf{k})
%  t^{\dagger}_{\alpha \mathbf{k} } t_{\alpha \mathbf{k}}
% + \frac{1}{2} \left( d_{\alpha\alpha}(\mathbf{k})
%  t_{\alpha \mathbf{k} }^{\dagger} t_{\alpha -\mathbf{k}}^{\dagger}
%  + h.c. \right) \label{Htri} \\
%& H_{\mu}^q(\mathbf{k}) = \bar{\epsilon}_{\mu\mu} (\mathbf{k})
%  q^{\dagger}_{\mu \mathbf{k} } q_{\mu \mathbf{k}}
% + \frac{1}{2} \left( \bar{d}_{\mu\mu }(\mathbf{k})
%  q_{\mu \mathbf{k} }^{\dagger} q_{\mu -\mathbf{k}}^{\dagger}
%  + h.c. \right),
\label{Hqui}
\end{align}
where $N$ is the system size. $\psi_{\mathbf{k}}^\dagger$ is 
the field operator of eight (triplet and quintet) excitonic boson operators with
momentum $\mathbf{k}$.  $\bm{\psi}^*_{-\mathbf{k}}$ is the transpose of
$\bm{\psi}^\dagger_{-\mathbf{k}}$.
Interaction matrices of triplet and quintet bosons
$\mathbf{h}(\mathbf{k})$ and $\mathbf{d}(\mathbf{k})$ are given as
\begin{subequations}
\begin{equation}
\mathbf{h}(\mathbf{k}) =
\begin{pmatrix} \epsilon_t-\epsilon_s & 0 \\
 0 & \epsilon_q-\epsilon_s \end{pmatrix}
+ \sum_\delta \begin{pmatrix} \mathbf{h}_\delta^{11} & \mathbf{h}_\delta^{12} \\
 \mathbf{h}_\delta^{21} & \mathbf{h}_\delta^{22}  \end{pmatrix}
e^{i\mathbf{k} \cdot \mathbf{r}_\delta},
\end{equation}
\begin{equation}
\mathbf{d}(\mathbf{k}) = \sum_\delta
\begin{pmatrix} \mathbf{d}_\delta^{11} & \mathbf{d}_\delta^{12} \\
 \mathbf{d}_\delta^{21} & \mathbf{d}_\delta^{22} \end{pmatrix}
e^{i\mathbf{k} \cdot \mathbf{r}_\delta}.
\end{equation}
\end{subequations}
The dispersions of the excitonic modes can be obtained by solving
Eq.~\ref{Hqui} with the help of
the Bogoliubov transformation (see appendix~\ref{Bogoliubov}).

%%%%%%%%%%%%%%%%%%%%%%%%%%%%%%%%%%%%%%%%%%%%%%%%%%%%%%%%%%%%%
\begin{figure}[t]
\centering
\includegraphics[width=\columnwidth]{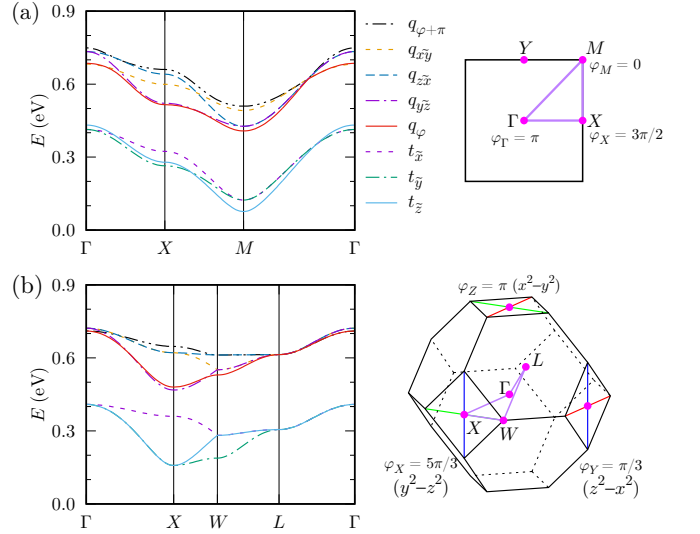}
\caption
{ (Color online)
  Dispersions of eight excitonic modes (triplets and quintets) and
  Brillouin zone (BZ) of (a) the square lattice
  and (b) the face-centered cubic lattice
  when $U=4.0$, $J_H=0.5$, $\lambda=0.4$, and $t=-0.4$ eV.
  Red, green, and blue lines in (b), which are parallel to $x$-, $y$-,
  and $z$-axes, refer to the minimum line of $t_x$, $t_y$, and $t_z$,
  respectively, when all couplings between triplet and quintet bosons
  are neglected.
  Eight excitonic modes are well separated into three triplet dominant modes
  ($t_{\tilde{x}},t_{\tilde{y}},t_{\tilde{z}}$),
  three $t_{2g}$-type quintet dominant modes
  ($q_{\tilde{xy}},q_{\tilde{yz}},q_{\tilde{zx}}$),
  and two $e_g$-type quintet modes ($q_{\varphi},q_{\varphi+\pi}$).
  $\varphi$ varies depending on $\mathbf{k}$.
  $\varphi$'s at high symmetric points are presented in the BZ.
}
\label{fig3}
\end{figure}
%%%%%%%%%%%%%%%%%%%%%%%%%%%%%%%%%%%%%%%%%%%%%%%%%%%%%%%%%%%%%

Figure~\ref{fig3} shows the dispersion of eight
excitonic modes on square and
fcc lattices with $U=4.0$, $J_H=0.5$, $\lambda=0.4$, and $t=-0.4$ eV.
As shown in the parameter tables presented in Appendix~\ref{inter_t},
all hopping matrices are well block diagonalized when eight bosons
are divided into four subsectors like
$\{ t_{x},q_{yz}\}$, $\{ t_{y},q_{zx}\}$, $\{ t_{z},q_{xy}\}$, and
$\{ q_{z^2},q_{x^2-y^2}\}$.
In addition, the energy splitting between triplet and quintet bosons
($\epsilon_q - \epsilon_t$) is considerably larger than
their coupling strengths. Eight excitonic modes are
well separated into three triplet dominant modes
($t_{\tilde{x}},t_{\tilde{y}},t_{\tilde{z}}$),
three $t_{2g}$-type quintet dominant modes
($q_{\tilde{xy}},q_{\tilde{yz}},q_{\tilde{zx}}$),
and two $e_g$-type quintet modes ($q_{\varphi},q_{\varphi+\pi}$).

On the square lattice, three triplet dominant modes show almost
the same dispersions. From $\Gamma$ through $X$ to $M$ points,
their energies monotonically decrease  (see Fig.~\ref{fig3}(a)).
The $t_{\tilde{z}}$ mode has maximum (minimum) at the $\Gamma$
($M$) point. Note that, the condensation of the $t_{\tilde{z}}$
bosons at $M$ point  may be interpreted as the antiferomagnetic
order with the $z$-component magnetic moment. The overall
dispersions for five quintet dominant modes are quite similar. The
$q_{\varphi_{\Gamma}+\pi}$ ($=q_{z^2}$) quintet has maximum value
at $\Gamma$ and $q_{\varphi_M}$ ($=q_{z^2}$) has a minimum at $M$
with $\varphi_{\Gamma}=\pi$ and $\varphi_{M}=0$.

On the fcc lattice, triplet dominant modes have a maximum value at  $\Gamma$,
whereas both $t_{\tilde{y}}$ and $t_{\tilde{z}}$ reach the minimum at $X$-point.
Their dispersions are consistent with those estimated in
recent theoretical studies~\cite{Pajskr2016,Chen2017}, in which
triplet states are only considered, except for their minimum pattern.
In contrast to our result, their minima appear on the $XW$ line away from  $X$-point
The discrepancy in the results stems from non-negligible coupling between triplet and quintet bosons, which was taken into account.
Our calculation also exhibits lines of minima which are highlighted with
red, green, and blue lines in Brillouin zone of Fig.~\ref{fig3}(b) when the coupling between triplet and quintet bosons is turned off.
In the presence of the coupling, % between triplet and quintet bosons,
dispersions of $t_{\tilde{y}}$ and $t_{\tilde{z}}$ modes at the $X$ point, to be shift down.
Minimum of the dispersions appear only at the $X$ (also $Y$ and $Z$) point.
Quintet dominant modes also show minimum values at  $X$, $Y$, and
$Z$ points.
As shown in Fig.~\ref{fig3}(b), $q_{\tilde{yz}}$ at $X$,
$q_{\tilde{zx}}$ at $Y$, and $q_{\tilde{xy}}$ at $Z$ have minimum energy.
Note that $\varphi_{\mathbf{k}}$'s at $X$, $Y$, and  $Z$ are $\frac{5\pi}{3}$, $\frac{\pi}{3}$, and
$\pi$, respectively, on the fcc lattice.
The quintet $q_{\varphi_X}$ ($=q_{y^2-z^2}$) excitation can be the lowest mode
when $U/t$ is small like Fig.~\ref{fig4}(b).

%%%%%%%%%%%%%%%%%%%%%%%%%%%%%%%%%%%%%%%%%%%%%%%%%%%%%%%%%%%%%
\begin{figure}[t]
\centering
\includegraphics[width=\columnwidth]{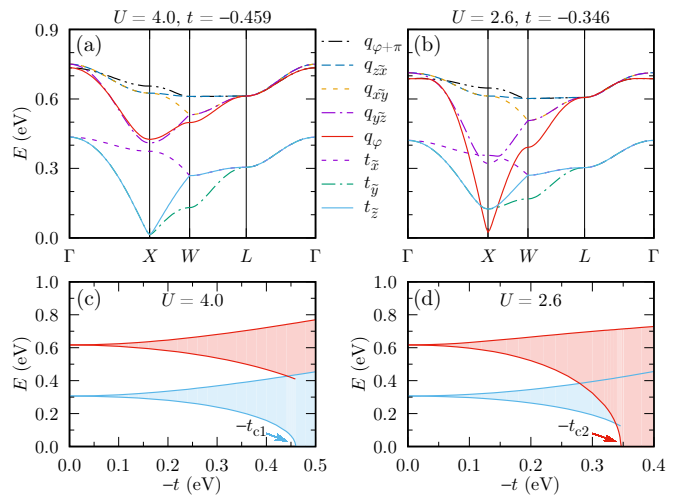}
\caption
{ (Color online)
  Dispersion of eight excitonic modes of the face-centered-cubic lattice
  when (a) $U=4.0$ and $t=-0.459$ eV and (b) $U=2.6$ and $t=-0.346$ eV.
  Excitation ranges of triplet and quintet dominant modes
  when (c) $U=4.0$ and (d) $U=2.6$ eV.  We set $J_H=0.5$ and $\lambda=0.4$ eV.
  Triplet and quintet condense when hopping strengths are larger than
  $-t_{c1}$ and $-t_{c2}$, respectively.
}
\label{fig4}
\end{figure}
%%%%%%%%%%%%%%%%%%%%%%%%%%%%%%%%%%%%%%%%%%%%%%%%%%%%%%%%%%%%%

One observes that
%the two types of bosons become increasingly dispersive with their minimum points
both triplet and quintet modes become more dispersive when the
hopping amplitude increases. Eventually, one of modes will
condense if the hopping strength exceeds a critical value which we
label $t_{c1}$ for the triplet and $t_{c2}$ for the quintet.
Because multiplet levels of triplets and quintets are about 0.31 and 0.62 eV, 
respectively, for $J_H=0.5$ and $\lambda=0.4$ eV,
%Because multiplet levels of quintets are about twice as high as 
%those of triplets,
it seems likely that the triplet dominant modes
always condenses first. For this reason only triplet bosons have
been took into account to explore the magnetic excitation of $d^4$
systems in previous works. Our calculation, however, shows that
such is not necessarily true for the fcc lattice.
Figure~\ref{fig4}(a) and (b) present dispersions of excitonic
modes
%triplet and quintet dispersions
on the fcc lattice for two different sets of parameters.
%When $U=4.0$ and $t=-0.5$ eV, minimum line of triplet is about to condense.
%The excitation of quintet is dispersive above $0.35$ eV.
When $U=4.0$ and $t=-0.459$ eV, $t_{\tilde{y}}$ and $t_{\tilde{z}}$ modes
are about to condense at the $X$ point.
Quintet dominant modes are dispersive above $0.4$ eV.
%When $U=2.8$ and $t=-0.4$ eV, in contrast, minimum point of quintet is close to zero
%even though spectra of triplet are located above $0.11$ eV.
%When $U=2.8$ and $t=-0.4$ eV, in contrast, minimum point of quintet is close to zero
%even though spectra of triplet are located above $0.11$ eV.
When $U=2.6$ and $t=-0.346$ eV, in contrast, minimum point of
$q_{\varphi=5\pi/3}$ ($q_{y^2-z^2}$) mode is close to zero at the $X$ point
even though spectra of triplet dominant modes are located above $0.1$ eV.
Figure~\ref{fig4}(c) and (d) give more detail how
triplet and quintet dominant modes are extended
as a function of hopping strength.
For a not too large $U$ of $2.6$ eV, even quintets of which the local excitation energy is
higher, soften very quickly with increasing hopping strength and condense first, in contrast to the situation for
a larger $U$ of $4.0$ eV, in which they exhibit similar a softening tendency as triplets are not the first instability of the system.

\section{Electronic excitation}

To get the dispersion of spin-orbit excitons, we assume that
the e-h excitations, which are described with
the partial density of state $\rho_{\rm eh}$ for two-sites,
appear in quite higher energy than spin-orbit excitons.
It is valid for the two-sites model for the considered parameters,
but it is not obvious for a lattice.
%The coordination number is four and twelve on square and
%fcc lattices, respectively.
When Coulomb repulsion is small enough in comparison to the
hopping parameter, the dispersion of e-h excitations can  overlap
with spin-orbit excitons. Eventually, charge gap closes and
insulator-metal transition (IMT) occurs. Then the effective
spin-orbital excitonic description breaks down. Therefore, it is
necessary to investigate the spectrum of e-h excitations as
functions of both hopping and Coulomb interaction.

%%%%%%%%%%%%%%%%%%%%%%%%%%%%%%%%%%%%%%%%%%%%%%%%%%%%%%%%%%%%%
\begin{figure}[t]
\centering
\includegraphics[width=\columnwidth]{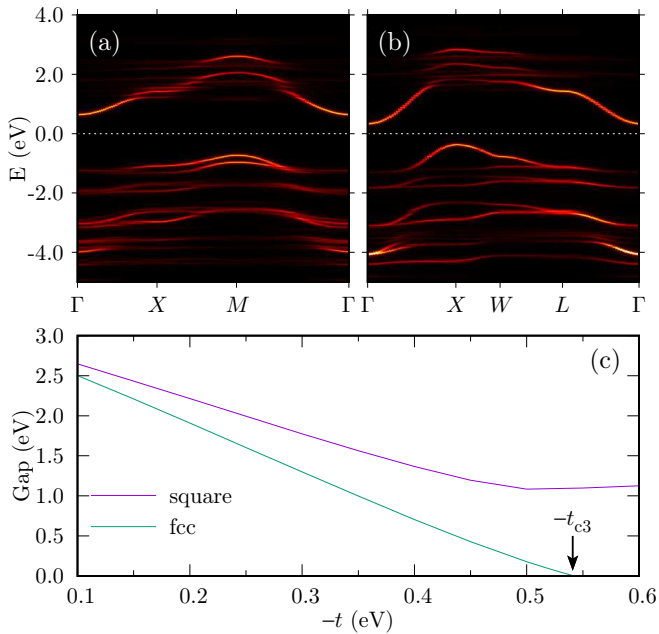}
\caption
{ (Color online)
  Spectral functions of (a) square lattice and
  (b) face-centered-cubic lattice calculated by variational CPT
  with four-site cluster
  when $U=4.0$, $J_H=0.5$, $\lambda=0.4$, and $t=-0.4$ eV.
  (c) The insulating gap behavior as a function of the hopping strength $t$
  when $U=4.0$, $J_H=0.5$, and $\lambda=0.4$ eV.
  When the hopping strength is larger than $-t_{c3}$,
  the insulating gap is closed and semimetallic phase is stabilized.
}
\label{fig5}
\end{figure}
%%%%%%%%%%%%%%%%%%%%%%%%%%%%%%%%%%%%%%%%%%%%%%%%%%%%%%%%%%%%%

We adopt the CPT~\cite{Senechal2002} with chemical potential optimization
via the variational cluster approximation (VCA)~\cite{Potthoff2003}.
We calculate the cluster Green's function
$\mathbf{G}'(z,\mu_c)$ with the four-site cluster presented
in inset of Fig.~\ref{fig1}(c) and (e). The cluster chemical potential $\mu_c$
is determined so that the average number per site is four in the ground state.
The lattice Green's function $\mathbf{G}(z,\mathbf{K})$ can be obtained
with $\mathbf{G}(z,\mathbf{K})^{-1}
=\mathbf{G}'(z,\mu)-(\mu-\mu_c)\mathbf{V}(\mathbf{K})$, where $\mu$ is
the lattice chemical potential and $\mathbf{V}(\mathbf{K})$ is
the Fourier transformation of the intercluster hopping matrix at
crystal momentum $\mathbf{K}$ of four-site supercell.
$\mu$ is optimized at the extremum point
of the grand functional $\Omega$ ($\frac{\partial \Omega}{\partial \mu}=0$)
\cite{Dahnken2004}.

Figures~\ref{fig5}(a) and (b) present spectral functions of
square and fcc lattice when $U=4.0$, $J_H=0.5$, $\lambda=0.4$ and $t=-0.4$ eV.
Both systems are located at the Mott insulating phase exhibiting
an indirect gap. When hopping strength increases, however,
the gap size decreases and IMT occurs at $t_{c3}$ point.
Due to the large coordination number,
the gap variation vs the hopping parameter on the fcc lattice
is more sensitive than that on the square lattice.
As shown in Fig.~\ref{fig5}(c), the gap on the fcc lattice is closed
even if the hopping strength is less than 0.6 eV
when $U=4.0$, $J_H=0.5$, and $\lambda=0.4$ eV, whereas
the gap is still finite on the square lattice.
Overall electronic phases for various parameters are presented in Fig.~\ref{fig6}.
Gray area in Fig.~\ref{fig6} refers to expected metallic region.

%%%%%%%%%%%%%%%%%%%%%%%%%%%%%%%%%%%%%%%%%%%%%%%%%%%%%%%%%%%%%
\begin{figure}[t]
\centering
\includegraphics[width=\columnwidth]{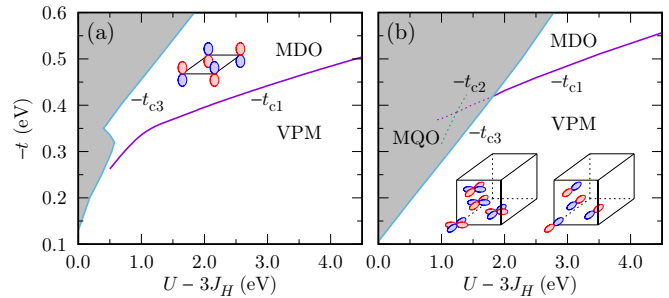}
\caption
{ (Color online)
  Phase diagram of (a) square lattice and (b) face-centered-cubic lattice.
  Triplet and quintet condensations give rise to the magnetic transitions
  from the Van Vleck paramagnetic (VPM) phase to the magnetic dipole order (MDO), and
  from the VPM phase to the magnetic quadrupole order (MQO), respectively.
  Insets in (a) and (b) refer to the schematic diagrams of
  possible condensed bosons in ordered phases.
  Gray area refers to expected region of the metallic phase based on
  the variational CPT calculation.
  The MQO phase on the fcc lattice could hardly emerge because of advanced
  charge gap closure.
}
\label{fig6}
\end{figure}
%%%%%%%%%%%%%%%%%%%%%%%%%%%%%%%%%%%%%%%%%%%%%%%%%%%%%%%%%%%%%

As shown in Fig.~\ref{fig6}(a), the spectral function of the square lattice
has an indirected gap determined at $\Gamma$ and $M$ points
in Mott insulating phase far from the IMT boundary.
However, its feature can vary in the vicinity of the IMT boundary 
when the hopping strength is large enough ($-t>0.35$ eV).
As shown in Fig.~\ref{fig7}(b), some portion of spectral weights transfers 
across the Fermi level around $\Gamma$ and $M$ points without the gap closure.
Shape of spectral function changes 
from Fig.~\ref{fig5}(a) to Fig.~\ref{fig7}(b) when $U$ decreasing.
However, the spectral weight transfer hardly occurs
when the hopping strength is small ($-t<0.32$ eV).
Overall shape of spectral function is almost robust until the gap is closed 
(see Fig.~\ref{fig7}(a)).
Consequently, the phase diagram of square lattice manifests different
types of phase boundary depending on the hopping strength and
shows the S-shape boundary around the intermediate hopping limit ($0.3 <-t< 0.4$ eV).
On the fcc lattice, in contrast, the gap is always closed indirectly 
at $\Gamma$ and $X$ points for given parameter range. 
The spectral weight transfer across the Fermi level is hardly manifested.
The phase boundary of IMT monotonically varies on the fcc lattice.

%%%%%%%%%%%%%%%%%%%%%%%%%%%%%%%%%%%%%%%%%%%%%%%%%%%%%%%%%%%%%
\begin{figure}[t]
\centering
\includegraphics[width=\columnwidth]{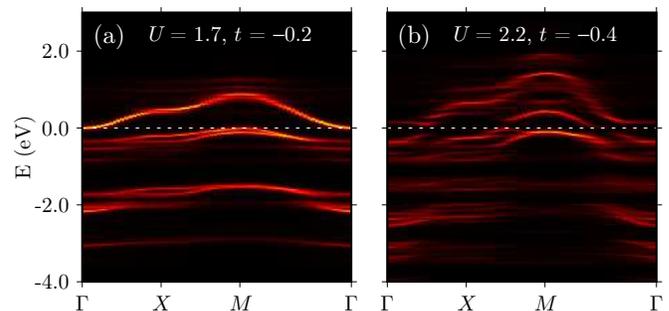}
\caption
{ (Color online)
  Spectral functions of square lattice calculated by variational CPT
  when (a) $U=1.7$, $J_H=0.5$, $\lambda=0.4$, and $t=-0.2$ eV,
  and (b) $U=2.2$, $J_H=0.5$, $\lambda=0.4$, and $t=-0.4$ eV.
% (c) The insulating gap and (d) occupied number 
% of $j_{\rm eff}=1/2$ and $3/2$ bands per site 
% ($n_{1/2}$ and $n_{3/2}$) as a function of $U-3J_H$
% when $t=-0.2$, $J_H=0.5$, and $\lambda=0.4$ eV (dark-violet line),
% and $t=-0.4$, $J_H=0.5$, and $\lambda=0.4$ eV (sea-green line).
}
\label{fig7}
\end{figure}
%%%%%%%%%%%%%%%%%%%%%%%%%%%%%%%%%%%%%%%%%%%%%%%%%%%%%%%%%%%%%

\section{Discussion}

Singlet, triplet, and quintet condensations  stabilize different
types of ground states, namely, Van Vleck paramagnetic (VPM) for
singlet,  magnetic dipole order (MDO) for triplet, and magnetic
quadrupole order (MQD) for quintet. The electronic and magnetic
phase diagram as functions of $t$ and $U-3J_H$ is shown in
Fig.~\ref{fig6}(a) and (b) on the square and fcc lattices,
respectively \cite{Hund}. On the square lattice the charge gap is
considerably larger than the excitonic gap in broad range of
parameters. With increase of the hopping parameter, the magnetic
ground state changes from VPM to MDO. The IMT may occur only in
weak $U-3J_H$ ($<0.5$ eV) limit. Because $t_{\tilde{z}}$ triplet
always condenses first, magnetic dipole moment is parallel to the
$z$-axis with the ($\pi,\pi$) ordering. The situation on the fcc
lattice is different. Because of large coordination number on the
fcc lattice, the dispersion of e-h excitations is much larger than
that on the square lattice. According to the magnetic phase
diagram obtained from the effective spin-orbital Hamiltonian of
two-site cluster, the instability towards condensation of
quintet modes of $q_{y^2-z^2}$, $q_{z^2-x^2}$, and $q_{x^2-y^2}$
 at $X$, $Y$, and $Z$ points, respectively, appears before that
towards the condensation of triplet dominant modes in weak
$U-3J_H$ ($<1.24$ eV) limit. It likely gives rise to the MQO like
the schematic diagram at lower part of Fig.~\ref{fig6}(b).
Considering the electronic excitation, however, the MQO could
hardly emerge on the fcc lattice. When $U-3J_H$ is less than about
1.82 eV, the VPM state can directly evolve to the metallic state
without intermediate  triplet or quintet condensations. Only in
large $U-3J_H$ ($>1.82$ eV) limit, the MDO can take place before
the IMT. In the phase, magnetic dipoles perpendicular to the $x$
($y$ or $z$) axis are parallelly ordered when they are in same
plane parallel to the $yz$ ($zx$ or $xy$) plane, whereas magnetic
dipoles in adjacent planes are antiparallely ordered.

The experimentally relevant values of $U-3J_H$ in iridates are
$0.4\sim2.0$ eV
\cite{Watanabe2010,Martins2011,Arita2012,BHKim2012,BHKim2014}. The
value the hopping parameter, for instance Ba$_2$YIrO$_6$ for which
band structure calculations find $t\sim 0.2$ eV~\cite{Dey2016}, is
still too low. The double-perovskite iridates are certainly
paramagnetic insulators with spin-orbital singlet in nature. Their
observed magnetic moment is surely attributed to not intrinsic but
extrinsic origins. Moreover, the phase diagram of the fcc lattice
implies that they barely exhibit any condensation of triplets or
quintets but show a ferromagnetic or paramagnetic metallic phase
in relevant parameter regime even enlarging the covalency and
weakening the on-site coulomb interaction. In this regard
materials, the typical situation for $4d$ transition metal
elements with small coordination number like Ca$_2$RuO$_4$ is
promising of the triplet condensation~\cite{Akbari2014,Jain2017}.

\section{Summary}

We investigated the spin-orbital excitations of pentavalent
iridium ions and their potential condensation on square and fcc
lattices. Based on the effective singlet-triplet-quintet
Hamiltonian, we have demonstrated that in both structures triplet
condensation can occur when the hopping $t$ becomes larger, for
still moderate values of the local electron-electron interaction
$U$. We have also demonstrated that in the fcc structure,depending
on the precise value of $U$, the condensation of quintets can be
the leading instability. Even though its local multiplet energy is
about double of triplets, a steep drop of the $q_{y^2-z^2}$
dispersion at $X$ point accelerates the softening. According to
the variational CPT calculation, however, the electronic charge
excitations are also strongly dispersive and the charge gap drops
much faster than those of spin-orbital excitations despite the
center of energy being even higher than those of  both triplets
and quintets. The magnetic transition driven by the condensation
of spin-orbital excitations is thus unlike to occur in iridium
double-perovskites such as Ba$_2$YIrO$_6$ because of strong SOC
strength and moderate Coulomb interaction -- $4d$ transition metal
compounds can be more promising in this respect.

\acknowledgements
 We thank Vamshi M. Katukuri for fruitful discussions.
 This work was supported by the DFG via SFB 1143, project A5.
 We also thank Korea Institute for Advanced Study for providing computing
  resources (KIAS Center for Advanced Computation Linux Cluster System).

\renewcommand{\thetable}{B\arabic{table}}
\setcounter{table}{0}

\appendix
\section{Bond-boson representation}
\label{append}

To describe lowest nine multiplets of $d^4$ in the strong SO limit,
we introduce nine boson operators.
Singlet boson ($J=0$) is
\begin{align}
& s^{\dagger} |vac\rangle =|T^0_0\rangle,
\end{align}
where $|vac\rangle$ is the vacuum state.
Triplet bosons ($J=1$) are
\begin{align}
& t_x^{\dagger} |vac\rangle =- \frac{1}{\sqrt{2}} \left(|T^{1}_{-1}\rangle-|T^{1}_{1}\rangle \right) \\
& t_y^{\dagger} |vac\rangle = -\frac{i}{\sqrt{2}} \left(T^{1}_{-1}\rangle+|T^{1}_{1}\rangle \right) \\
& t_z^{\dagger} |vac\rangle = |T^{1}_{0}\rangle.
\end{align}
Quintet bosons ($J=2)$ are
\begin{align}
& q_{xy}^{\dagger} |vac\rangle = \frac{i}{\sqrt{2}} \left(|T^{2}_{-2}\rangle-|T^{2}_{2}\rangle \right) \\
& q_{yz}^{\dagger} |vac\rangle = \frac{i}{\sqrt{2}} \left(|T^{2}_{-1}\rangle+|T^{2}_{1}\rangle \right) \\
& q_{z^2}^{\dagger} |vac\rangle = |T^{2}_{0}\rangle \\
& q_{zx}^{\dagger} |vac\rangle = \frac{1}{\sqrt{2}} \left(|T^{2}_{-1}\rangle-|T^{2}_{1}\rangle \right) \\
& q_{x^2-y^2}^{\dagger} |vac\rangle = \frac{1}{\sqrt{2}} \left(|T^{2}_{-2}\rangle+|T^{2}_{2}\rangle \right).
\end{align}
In cubic or tetragonal system, it is more convenient to use new coordinate to
describe quintet bosons given by $q_{xy}^{\dagger}$, $q_{yz}^{\dagger}$,
$q_{zx}^{\dagger}$, $q_{\varphi}^{\dagger}$, and $q_{\varphi+\pi}^{\dagger}$ where
\begin{align}
& q_{\varphi}^{\dagger} = \cos\frac{\varphi}{2} q_{z^2}^{\dagger}
   - \sin\frac{\varphi}{2} q_{x^2-y^2}^{\dagger}.
\label{varphi}
\end{align}
$q_{\varphi}^{\dagger}$ can be $-q_{z^2}^{\dagger}$ for $\varphi=0$,
$q_{x^2}^{\dagger}$ for $2\pi/3$, $q_{y^2}^{\dagger}$ for $4\pi/3$,
$-q_{x^2-y^2}^{\dagger}$ for $\pi$, $q_{z^2-x^2}^{\dagger}$ for $\pi/3$,
or $q_{y^2-z^2}^{\dagger}$ for $5\pi/3$.

\section{Interaction strength}
\label{inter_t}

\begin{table}[H]
\caption{
Non-zero interaction parameters between triplet bosons
of the square and face-center-cubic (fcc) lattices
when $U=4.0$, $J_H=0.5$, $\lambda=0.4$, and $t=-0.4$ eV.
On the square (fcc) lattice, triplet indices $\{1,2,3\}$ represent
$\{y,z,x\}$ along the $x$-axis ($yz$-plane), $\{z,x,y\}$ along the $y$-axis ($zx$-plane),
or $\{x,y,z\}$ along the $z$-axis ($xy$-plane).
Unit is eV.
}
\label{TB1}
\begin{ruledtabular}
\begin{tabular}{c c c | c c c}
 & square & fcc & & square & fcc \\
\hline
  $[h_\delta^{11}]_{11}$ & $0.0404$ & $0.0040$ &
  $[d_\delta^{11}]_{11}$ & $0.0226$ & $0.0041$ \\
  $[h_\delta^{11}]_{22}$ & $0.0404$ & $0.0040$ &
  $[d_\delta^{11}]_{22}$ & $0.0226$ & $0.0041$ \\
  $[h_\delta^{11}]_{33}$ & $0.0256$ & $0.0231$ &
  $[d_\delta^{11}]_{33}$ & $0.0295$ & $0.0248$
\end{tabular}
\end{ruledtabular}
\end{table}

\begin{table}[H]
\caption{
Non-zero interaction parameters between quintet bosons of
the square and face-center-cubic (fcc) lattices
when $U=4.0$, $J_H=0.5$, $\lambda=0.4$, and $t=-0.4$ eV.
On the square (fcc) lattice, quintet indices
$\{4,5,6,7,8\}$ represent
$\{yz,zx,x^2,xy,y^2-z^2\}$ along the $x$-axis ($yz$-plane),
$\{zx,xy,y^2,yz,z^2-x^2\}$ along the $y$-axis ($zx$-plane),
or $\{xy,yz,z^2,zx,x^2-y^2\}$ along the $z$-axis ($xy$-plane).
Unit is eV.
}
\label{TB2}
\begin{ruledtabular}
\begin{tabular}{c c c | c c c}
 & square & fcc & & square & fcc \\
\hline
  $[h_\delta^{22}]_{44}$ & $0.0530$ & $0.0003$ &
  $[d_\delta^{22}]_{44}$ & $-0.0017$ & $~\ 0.0006$ \\
  $[h_\delta^{22}]_{55}$ & $0.0233$ & $0.0153$ &
  $[d_\delta^{22}]_{55}$ & $-0.0294$ & $-0.0188$ \\
  $[h_\delta^{22}]_{66}$ & $0.0116$ & $0.0186$ &
  $[d_\delta^{22}]_{66}$ & $~\ 0.0118$ & $~\ 0.0250$ \\
  $[h_\delta^{22}]_{77}$ & $0.0233$ & $0.0153$ &
  $[d_\delta^{22}]_{77}$ & $-0.0294$ & $-0.0188$ \\
  $[h_\delta^{22}]_{88}$ & $0.0530$ & $0.0003$ &
  $[d_\delta^{22}]_{88}$ & $-0.0017$ & $~\ 0.0006$
\end{tabular}
\end{ruledtabular}
\end{table}

\begin{table}[H]
\caption{
Non-zero interaction parameters between triplet and quintet bosons of
the square and face-center-cubic (fcc) lattices
when $U=4.0$, $J_H=0.5$, $\lambda=0.4$, and $t=-0.4$ eV.
Indices 1 -- 8 are same as those presented in table~\ref{TB1} and~\ref{TB2}.
Unit is eV.
}
\label{TB3}
\begin{ruledtabular}
\begin{tabular}{c c c | c c c}
 & square & fcc & & square & fcc \\
\hline
  $[h_\delta^{12}]_{15}$ & $~\ 0.0029i$ & $-0.0088i$&
  $[d_\delta^{12}]_{15}$ & $~\ 0.0035i$ & $-0.0086i$ \\
  $[h_\delta^{12}]_{27}$ & $-0.0029i$   & $~\ 0.0088i$&
  $[d_\delta^{12}]_{27}$ & $~\ 0.0035i$ & $~\ 0.0086i$ \\
  $[h_\delta^{21}]_{51}$ & $-0.0029i$   & $~\ 0.0088i$&
  $[d_\delta^{21}]_{51}$ & $-0.0035i$   & $-0.0086i$ \\
  $[h_\delta^{21}]_{72}$ & $~\ 0.0029i$ & $-0.0088i$&
  $[d_\delta^{21}]_{72}$ & $-0.0035i$   & $~\ 0.0086i$
\end{tabular}
\end{ruledtabular}
\end{table}

\section{Mean-field Hamiltonian}
\label{meanH}

Because of the hardcore condition, only one boson will be occupied at each site
with following relation
\begin{equation}
s_i^{\dagger}s_i+\sum_{\alpha} t_{\alpha i}^{\dagger} t_{\alpha i}
 + \sum_{\mu} q_{\mu i}^{\dagger} q_{\mu i} = 1.
\end{equation}
We have to solve following Hamiltonian:
\begin{equation}
H_{\rm eff} + \sum_i \lambda_L \Big( s_i^{\dagger}s_i+
 \sum_{\alpha} t_{\alpha i}^{\dagger} t_{\alpha i}
 + \sum_{\mu} q_{\mu i}^{\dagger} q_{\mu i} -1 \Big).
\end{equation}
Provided that the ground state in which singlet bosons almost condense,
we can treat $s_i (s_i^{\dagger})$ as scalar value $s$ according
to the mean-field theory of Bose system.
Let ${\bm \psi}_i^\dagger $ be the field operator of
eight (triplet and quintet) excitonic bosons at the $i$-th site like
\begin{equation}
{\bm \psi}_i^\dagger =
\begin{pmatrix}
t_{x,i}^\dagger  & t_{y,i}^\dagger & t_{z,i}^\dagger &
q_{xy,i}^\dagger & q_{yz,i}^\dagger & q_{zx,i}^\dagger &
q_{z^2,i}^\dagger & q_{x^2-y^2,i}^\dagger
\end{pmatrix}.
\end{equation}
We can get following the mean-field Hamiltonian from Eq.~\ref{Heff}:
\begin{align}
& H_{MF}^{s,\lambda_L} \approx N(\epsilon_s s^2+ \lambda_L s^2-\lambda_L)
\nonumber \\
&~ + \frac{1}{2} \sum_{i\delta} \left[ \bm{\psi}_{i_\delta}^\dagger
 \mathbf{h}^{s,\lambda_L}_\delta \bm{\psi}_i +
  \bm{\psi}_{i_\delta}^\dagger \mathbf{d}_{\delta}^{s,\lambda_L}
       \bm{\psi}_i^* + h.c. \right]
\nonumber \\
&~ + \cdots,
\end{align}
where
\begin{subequations}
\begin{equation}
\mathbf{h}^{s,\lambda_L}_\delta =
\begin{pmatrix} \epsilon_t+\lambda_L & 0 \\

 0 & \epsilon_q+\lambda_L \end{pmatrix}
+ s^2 \begin{pmatrix} \mathbf{h}_\delta^{11} & \mathbf{h}_\delta^{12} \\
 \mathbf{h}_\delta^{21} & \mathbf{h}_\delta^{22}  \end{pmatrix},
\end{equation}
\begin{equation}
\mathbf{d}^{s,\lambda_L}_\delta = s^2
\begin{pmatrix} \mathbf{d}_\delta^{11} & \mathbf{d}_\delta^{12} \\
 \mathbf{d}_\delta^{21} & \mathbf{d}_\delta^{22} \end{pmatrix},
\end{equation}
\end{subequations}
and $N$ is total number of sites.
$\bm{\psi}_i$ and $\bm{\psi}_i^*$ are the Hermitian conjugate
and transpose of $\bm{\psi}_i^\dagger$ vector, respectively.
%based on following Fourier transformation relations:
%\begin{align}
%&\epsilon_{\alpha\beta}^{s,\lambda_L}(\mathbf{k})=(\epsilon_t+\lambda_L)\delta_{\alpha\beta}
% +s^2 \sum_\delta h_{\alpha\beta}^\delta e^{i\mathbf{k} \cdot \mathbf{r}_\delta},\\
%& d_{\alpha\beta}^s (\mathbf{k})= s^2 \sum_\delta d_{\alpha\beta}^\delta
% e^{i\mathbf{k} \cdot \mathbf{r}_\delta},\\
%&\bar{\epsilon}_{\mu\nu}^{s,\lambda_L} (\mathbf{k}) =(\epsilon_q+\lambda_L)\delta_{\mu\nu}
% +s^2 \sum_\delta \bar{h}_{\mu\nu}^\delta e^{i\mathbf{k} \cdot \mathbf{r}_\delta},\\
%& \bar{d}_{\mu\nu}^s (\mathbf{k}) = s^2 \sum_\delta \bar{d}_{\mu\nu}^\delta
% e^{i\mathbf{k} \cdot \mathbf{r}_\delta},
%\end{align}
%where $\mathbf{r}_\delta$ is the displacement vector between Ir's along the $\delta$-direction.
%Because the mean-field order parameter $s^2$ follows the saddle-conditions
%$\partial H_{MF}^{s,\lambda_L}/\partial s^2 = 0$, we can get $\lambda_L=-\epsilon_s$.
%Also, we can approximate $s^2 = 1$ when there is no condensation of a triplet or
%quintet boson.
%Finally, we can get following Hamiltonian:
Because the mean-field order parameter $s^2$ follows the saddle-conditions
$\partial H_{MF}^{s,\lambda_L}/\partial s^2 = 0$,
we can get $\lambda_L=-\epsilon_s$ and approximate $s^2 = 1$
when there is no condensation of a triplet or quintet boson.
Finally, we can get following Hamiltonian:
\begin{align}
& H_{MF} \approx N\epsilon_s +
\nonumber \\
&~+ \sum_{\mathbf{k}} \left[
   \bm{\psi}^\dagger_\mathbf{k} \mathbf{h} (\mathbf{k})
   \bm{\psi}_\mathbf{k}+
   \frac{1}{2} \left(
   \bm{\psi}^\dagger_\mathbf{k} \mathbf{d} (\mathbf{k})
   \bm{\psi}^*_{-\mathbf{k}}+  h.c.
   \right)
\right] \nonumber \\
&~ + \cdots
\end{align}
based on following Fourier transformation relations:
\begin{subequations}
\begin{equation}
\bm{\psi}_\mathbf{k} = \frac{1}{\sqrt{N}}
\sum_i \bm{\psi}_i e^{i\mathbf{k} \cdot \mathbf{r}_i},
\end{equation}
\begin{equation}
\mathbf{h}(\mathbf{k}) =
\begin{pmatrix} \epsilon_t-\epsilon_s & 0 \\
 0 & \epsilon_q-\epsilon_s \end{pmatrix}
+ \sum_\delta \begin{pmatrix} \mathbf{h}_\delta^{11} & \mathbf{h}_\delta^{12} \\
 \mathbf{h}_\delta^{21} & \mathbf{h}_\delta^{22}  \end{pmatrix}
e^{i\mathbf{k} \cdot \mathbf{r}_\delta},
\end{equation}
\begin{equation}
\mathbf{d}(\mathbf{k}) = \sum_\delta
\begin{pmatrix} \mathbf{d}_\delta^{11} & \mathbf{d}_\delta^{12} \\
 \mathbf{d}_\delta^{21} & \mathbf{d}_\delta^{22} \end{pmatrix}
e^{i\mathbf{k} \cdot \mathbf{r}_\delta},
\end{equation}
\end{subequations}
where $\mathbf{r}_i$ and $\mathbf{r}_\delta$ are the position vector of an $i$-site and
the displacement vector between Ir's along the $\delta$-direction,
respectively.

\section{Bogoliubov transformation}
\label{Bogoliubov}

Because of the matrix $\mathbf{d}(\mathbf{k})$, the field operator
$\psi_{\mathbf{k}}^\dagger$ is coupled with $\psi_{-\mathbf{k}}^\dagger$.
The mean-field Hamiltonian written in Eq.~\ref{Hqui} can be extended
as following
\begin{align}
& H_{MF} = N\epsilon_s -\frac{1}{2}
\sum_{\mathbf{k}} \textrm{Tr}\left[ \mathbf{h}(\mathbf{k}) \right] \nonumber \\
&~ + \frac{1}{2} \sum_{\mathbf{k}}
\begin{pmatrix}
\bm{\psi}_\mathbf{k}^\dagger & \bm{\psi}_{-\mathbf{k}}^\intercal
\end{pmatrix}
\begin{pmatrix}
\mathbf{h}(\mathbf{k}) &  \mathbf{d}(\mathbf{k}) \\
\mathbf{d}(\mathbf{k})^\dagger  &  \mathbf{h}(-\mathbf{k})^\intercal
\end{pmatrix}
\begin{pmatrix}
\bm{\psi}_\mathbf{k} \\ \bm{\psi}^*_{-\mathbf{k}}
\end{pmatrix},
\end{align}
where $\bm{\psi}_{-\mathbf{k}}^\intercal$ and
$\mathbf{h}(-\mathbf{k})^\intercal$ are the transpose of
$\bm{\psi}_{-\mathbf{k}}$ and $\mathbf{h}(-\mathbf{k})$, respectively.
We can get the dynamic equation of motion of both
$\bm{\psi}_\mathbf{k}$ and $\bm{\psi}_{-\mathbf{k}}^\dagger$
operators as following
\begin{equation}
i \hbar \partial_t
\begin{pmatrix}
\bm{\psi}_\mathbf{k} \\
\bm{\psi}_{-\mathbf{k}}^*
\end{pmatrix}
=
\begin{pmatrix}
\mathbf{h}(\mathbf{k}) &  \mathbf{\Delta}(\mathbf{k}) \\
-\mathbf{\Delta}(\mathbf{k})^\dagger  & -\mathbf{h}(-\mathbf{k})^\intercal
\end{pmatrix}
\begin{pmatrix}
\bm{\psi}_\mathbf{k} \\
\bm{\psi}_{-\mathbf{k}}^*
\end{pmatrix},
\label{Eq_motion}
\end{equation}
where $\mathbf{\Delta}(\mathbf{k})
=\frac{1}{2}\left(\mathbf{d}(\mathbf{k})
   +\mathbf{d}(-\mathbf{k})^\intercal\right)$~\cite{Maldonado1993}.
By solving the general eigenvalue problem of Eq.~\ref{Eq_motion},
we can get dispersions of excitonic normal modes.

\section{Magnetization}
\label{mag}

In terms of bond-boson operators, local magnetic moment along the $\gamma$-direction
is expressed by following relation:
\begin{multline}
 M_{\gamma} = A \left(t_{\gamma}^{\dagger}s+s^{\dagger}t_{\gamma}\right) +
 \tilde{g}J_{\gamma}  \\
 + \sum_{\alpha\beta} \left( m_{\alpha\beta}^{\gamma} t_{\alpha}^{\dagger} q_{\beta}
 +h.c.\right),
\end{multline}
where $A$ is constant and $\tilde{g}$ is the $g$-tensor expressed into
$\tilde{g}=g_0 \oplus g_1 \oplus g_2$ where $g_j$ is the $g$-tensor of a $J=j$ state.
The matrix $\mathbf{m}^{z}$ has only three nonzero elements such as
$m_{x,zx}^z=m_{y,yz}^{z}$ and $m_{z,z^2}^{z}$.
When $J_H>>\lambda$, $A=\sqrt{6}$, $g_0=0$, and $g_1=g_2=1/2$.
Besides, $A\approx 2.41$, $g_0=0$, $g_1=1/2$, and $g_2 \approx 0.36$
when $J_H=0.5$ and $\lambda=0.4$ eV.

\end{document}